\documentclass[prl,twocolumn,aps,showpacs]{revtex4}
\usepackage{graphicx}
\usepackage{epsfig}
\usepackage{bm,hyperref}

\def\be{\begin{equation}}
\def\ee{\end{equation}}
\def\ba{\begin{eqnarray}}
\def\ea{\end{eqnarray}}

\begin{document}
\title{Two Critical velocities for a superfluid in a periodic potential}
\author{Biao Wu}
\affiliation{Institute of Physics and ICQS, Chinese Academy of Sciences,
Beijing 100080, China}
\author{Junren Shi}
\affiliation{Institute of Physics and ICQS, Chinese Academy of Sciences,
Beijing 100080, China}
\date{\today}
%
\begin{abstract}
  In contrast to a homogeneous superfluid which has only one critical
  velocity, there exist two critical velocities for a superfluid in a
  periodic potential. The first one, which we call inside critical
  velocity, is for a macroscopic impurity to move frictionlessly in
  the periodic superfluid system; the second, which is called trawler
  critical velocity, is the largest velocity of the lattice for the
  superfluidity to maintain.  The result is relevant to the
  superfluidity observed in the Bose-Einstein condensate in an optical
  lattice and supersolid helium.
\end{abstract}
\pacs{67.80.-s,67.40.-w,03.75.Lm}
\maketitle
One characteristic feature of superfluidity is the existence of
critical velocity. According to Landau's theory of
superfluidity\cite{Landau1941USSR}, this critical velocity is given by
the speed of sound. In developing his theory, Landau had his focus on
a homogeneous superfluid, which has continued to be the focus of later
studies\cite{Landau_SM2,Nozieres1990Book}. Experimental advances in
recent years have brought to people's attention a different type of
systems, superfluids in periodic potentials. They include a
Bose-Einstein condensate (BEC) in an optical
lattice\cite{Morsch2006RMP} and supersolid helium\cite{Kim2004Sci}. In the
BEC system, the lattice is created by counter-propagating laser beams.
For supersolid helium, the superfluid is defects (vacancies or
interstitials) in the lattice self-assembled by helium
atoms as commonly believed\cite{Andreev1969JETP,Leggett1970PRL}. 
Interestingly, the crust of a neutron star can also be considered as 
a superfluid in a lattice\cite{Lattimer2004Sci,Carter2005NPA}.
In addition, it is expected that the superfluidity of a paired Fermi gas 
in an optical lattice will be a subject of intensive 
investigation\cite{Ketterle2005Nature}.

\begin{figure}[!htb]
\includegraphics[width=7.0cm]{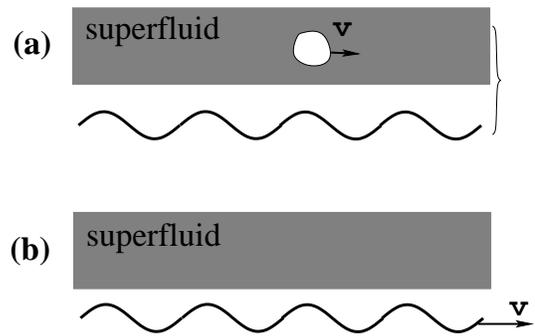}
\caption{(a) A macroscopic obstacle moves with a  velocity of ${\bf v}$
inside a superfluid residing in a periodic potential. The curly
brace indicates that the superfluid and the periodic potential
are ``locked'' together and there is no relative motion between them.
(b) The lattice where a superfluid resides is slowly accelerated 
to a velocity of ${\bf v}$.}
\label{fig:slatt}
\end{figure}

In this Letter we show that the presence of the periodic potential has
non-trivial consequences, requiring a revisit of the concept
of  critical velocity.  In contrast to the homogeneous
superfluid which has only one critical velocity, there are two
distinct critical velocities for a superfluid in a periodic potential.
The first one, which we call inside critical velocity, is for an
impurity to move frictionlessly in the periodic superfluid
system (Fig.\ref{fig:slatt}(a)); the second, which is called trawler
critical velocity, is the largest velocity of the lattice 
for the superfluidity to maintain (Fig.\ref{fig:slatt}(b)).  
These two  critical velocities will be illustrated with
a BEC in a one-dimensional optical lattice.

The presence of the periodic potential plays a decisive role in the
appearance of the two critical velocities. Because of the addition of
a periodic potential, two very different situations can arise in the
superfluid system. The first situation is described in
Fig.\ref{fig:slatt}(a), where one macroscopic impurity moves inside
the superfluid. The key feature in this situation is that there is no
relative motion between the superfluid and the periodic potential.
Fig.\ref{fig:slatt}(b) illustrates the other situation, where the
lattice is slowly accelerated to a given velocity and there is a
relative motion between the superfluid and the periodic potential.
For these two different situations, naturally arise two critical
velocities.
 

Both critical velocities can be measured with BECs in
optical lattices. The inside critical velocity $v_i$ can be measured
with the same experimental setting as in Ref.\cite{Onofrio2000PRL},
where the superfluidity of a BEC was studied by moving a blue-detuned
laser inside the BEC.  For the trawler critical velocity $v_t$, one
can repeat the experiment in Ref.\cite{Fallani2004PRL} where a BEC is
loaded in a moving optical lattice. One only needs to shift his
attention from dynamical instability to superfluidity. For solid
helium, only the trawler critical velocity can be measured.

\begin{figure}[!htb]
\includegraphics[width=7.0cm]{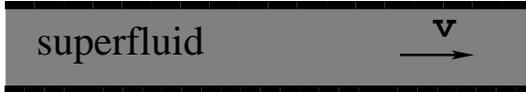}
\caption{A superfluid moves inside a small
tube with a velocity of ${\bf v}$.}
\label{fig:stube}
\end{figure}

It is instructive to briefly review Landau's theory of critical 
velocity for superfluid
before further discussion\cite{Landau_SM2}. 
Consider a superfluid moving inside a small tube with a velocity of $\bf v$
as in Fig.\ref{fig:stube}. Suppose a single elementary excitation is 
generated and its energy 
is $\varepsilon_0(\bf p)$ and momentum is $\bf p$ in the superfluid frame 
where the superfluid is at rest, then by the usual Galilean transformation,
we have the excitation energy:
\be
\varepsilon({\bf v},{\bf p})=\varepsilon_0({\bf p})-{\bf v}\cdot{\bf p}\,,
\ee
in the reference frame where the tube is motionless. Landau argues
that if $\varepsilon({\bf v},{\bf p})$ is negative for some excitations,
these excitations are energetically favored to be generated and,
therefore, the superfluidity is unstable. Since the low energy excitations
are phonons $\varepsilon_0({\bf p})=up$, it is clear that 
$\varepsilon({\bf v},{\bf p})$
can be negative only when $v>u$. This implies that the sound speed $u$
is the critical velocity beyond which the superfluid becomes normal fluid. 

It is clear that the ``tube'' frame holds a special position in
Landau's theory. It is only in this frame the tube does not disturb the
system so that the normal fluid consisting of the excitation ``gases''
can be in a thermal equilibrium with the
environment~\cite{Pitaevskii2003Book}. However, in the presence of the
periodic potential, the disturbance can be from either the external
impurity (Fig.~\ref{fig:slatt}(a)) or the imperfection of the periodic
potential (Fig.~\ref{fig:slatt}(b)), and one needs to define the
different ``tube'' frames for the different scenarios.  For this
reason and convenience of our future discussion, we call such a frame as
thermodynamics frame.


We focus first on the situation of Fig.\ref{fig:slatt}(a), where a
macroscopic obstacle moves inside the periodic superfluid with a
velocity $\bf{v}$.  We follow closely Landau's
argument\cite{Landau_SM2} and start our discussion in the superfluid
frame where the superfluid is at rest.  In such a periodic system, the
elementary excitation (quasi-particle) is characterized by
quasi-momentum ${\bf q}$ and the Bloch-band index $n$.  If the excited
state is described by the wavefunction $\Psi_{n\bf q}$, then its energy
and momentum are, respectively, 
\ba 
\varepsilon_n({\bf
  q})&=&\langle\Psi_{n\bf q}|\hat{H}|\Psi_{n\bf q}\rangle-E_0
\,,\label{excitation}\\
{\bf p}_n({\bf q})&=&\langle\Psi_{n\bf q}|\hat{\bf p}|\Psi_{n\bf
  q}\rangle\,, 
\ea 
where $\hat{H}$ is the Hamiltonian of system with $E_0$ being the
ground state energy.  $\hat{\bf p}=\sum_j{\bf p}_j$ is the total
momentum operator of the system.  Note that without the periodic
potential one would have simply ${\bf p}=\hbar{\bf q}$. With the
periodic potential, this simple relation no longer holds. In the
thermodynamic frame in which the obstacle is motionless, the
Hamiltonian of the system is transformed to: 
\be \hat{H}^\prime =
\hat{H} - \mathbf{v}\cdot \hat{\bf p} + \frac{1}{2}M\mathbf{v}^2 
\ee where
$M$ is the total mass of the system.  The excitation energy in this
frame reads, 
\be 
\varepsilon^\prime_n({\bf v}, {\bf q})=
\langle\Psi_{n\bf q}|\hat{H}^\prime|\Psi_{n\bf
  q}\rangle-E^\prime_0= \varepsilon_n({\bf q})-{\bf v}\cdot{\bf p}_{\it n}(\bf
q)\,, 
\ee 
where $E_0^\prime=E_0+Mv^2/2$ is the ground state energy in the
thermodynamics frame.  This excitation energy determines the stability
of the system.  If it is positive for all values of $\bf q$ and band
index $n$, then the excitation of quasi-particles is not energetically
favored and the system is a superfluid.  Otherwise, the
quasi-particles can be generated spontaneously and the liquid flow
experiences viscosity. We can thus define a
critical velocity, beyond which $\varepsilon_n^\prime(\bf q)$ can be negative for
some values of $n,\,\bf q$. 
This critical velocity $v_i$ is given by 
\be v_i={\rm
  Minimum~of~}\frac{\varepsilon_n({\bf q})}{|\bf{p}_{\it n}(\bf q)|}\,.  
\ee
We call $v_i$ inside critical velocity to distinguish it from the
other critical velocity that we shall discuss next.  

\begin{figure}[!htb]
\includegraphics[width=7.5cm]{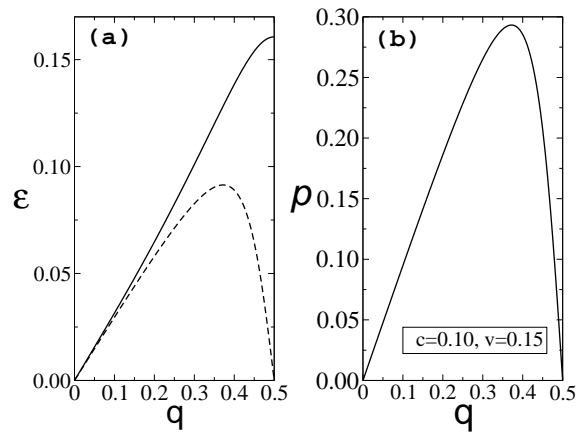}
\caption{(a) Energy $\varepsilon_0(q)$ and (b) momentum $p(q)$ of a
  quasi-particle for a BEC in a one-dimensional optical lattice. The
  dashed line in (a) is for $up(q)$ with $u$ being the sound speed.}
\label{fig:peng}
\end{figure}

We take a look at an example, a BEC in a one-dimensional optical
lattice. We treat such a system with the standard Gross-Pitavskii
equation,
\be
\label{eq:gp}
i\frac{\partial}{\partial t}\psi=-\frac{1}{2}\frac{\partial^2}{\partial x^2}\psi
+v(\cos x)\psi+c|\psi|^2\psi\,. 
\ee
In the above equation, the energy is in units of $4\pi^2\hbar^2m/a^2$,
the unit of the momentum is $2\pi\hbar/a$, and $q$ is measured in 
units of $\pi/a$ ($a$ is the period of the optical lattice). 
For the details of the unit system, please refer to Ref.\cite{Wu2003NJP}. 
Following the procedure in Ref.\cite{Wu2003NJP}, we can compute its Bogoliubov 
excitations and, consequently, the energies and momenta of
these excitations. Plotted in Fig.\ref{fig:peng} are the energy and momentum of the
quasi-particle in this simple system.  We see from this figure that
the dashed line for $u p(q)$ lies completely under $\varepsilon(q)$. This
means that the critical velocity $v_i$ is exactly the sound speed $u$
in this simple example. 


The motion of an impurity inside a superfluid has been studied for
a long time by moving ions inside liquid helium\cite{Fetter1976InBook}.
This technique was used to verify Landau's original criterion of 
superfluidity\cite{Reif1960PR}, which is impossible to verify by 
flowing superfluid helium inside a small tube as one would generate vortex 
and turbulence and destroy superfluidity well before the superflow 
reaches the sound speed\cite{Nozieres1990Book}.
More recently, an obstacle created by a blue-detuned laser beam
was moved back and forth inside a rather homogeneous BEC
to test its superfluidity\cite{Onofrio2000PRL}. We believe
that this technique can also  be used to measure the inside critical 
velocity for a BEC in an optical lattice. On the theoretical side,
the motion of an impurity in a homogeneous superfluid was
studied by Girardeau, who found the ``onset of acoustical wave
drag as the impurity speed reaches the speed of sound''\cite{Girardeau1961PoF}.

We study now the case depicted in Fig.\ref{fig:slatt}(b), where the
lattice is accelerated slowly to a given velocity $\bf v$.  In this
case, the thermodynamics frame is the frame that moves along with the
lattice. Viewing from the thermodynamics frame, the slow acceleration
induces an adiabatic evolution of the Bose-Einstein condensation from
the Bloch state at $\Gamma$ point (${\bf k} = 0$) to a Bloch state
with non-vanishing Bloch wave vector ${\bf k}=-m{\bf
  v}/\hbar$~\cite{ChoiAndNiu1999,Morsch2001PRL}.  The same effect can
be achieved by slowly turning on a moving
lattice\cite{Fallani2004PRL}.
The excitation energy in the thermodynamics frame can be obtained in a
similar way as that in Eq.~(\ref{excitation}), albeit the ``ground state''
for the present case has a macroscopic condensation at nonzero 
${\bf k}$. We thus have, 
\be 
\tilde{\varepsilon}^\prime_n({\bf v}, {\bf
  q})=\langle\Psi_{n{\bf q};{\bf k}}|\hat{H}|\Psi_{n{\bf q};{\bf k}}
\rangle-\tilde{E}_0({\bf k}) \,, 
\label{trawlerexcitation}
\ee 
where $\tilde{E}_0({\bf k})$ is
the energy of the ground state and $|\Psi_{n{\bf
    q};{\bf k}} \rangle$ denotes the excitation state with
quasi-momentum ${\bf q}$.  We note that in the present case both the ground
state and the excitation state have dependence on
the condensation momentum ${\bf k}\equiv -m{\bf v}/\hbar$.
As a result, 
the excitation energy $\tilde{\varepsilon}_n^\prime({\bf v},{\bf q})$ depends
on both the velocity $\bf v$ of the periodic potential and the
quasi-momentum $\bf q$ of the excitation.  The stability of the
superfluid phase can be determined in the same spirit as that in the
first case: if $\tilde{\varepsilon}_n^\prime({\bf v},{\bf q})$ can be negative for
some finite $\bf q$, then the superfluidity is lost. The critical
velocity for this situation is given by the smallest $v$ such that
$\tilde{\varepsilon}_n^\prime({\bf v},{\bf q})=0$ for some finite $\bf q$. 
 We denote
it as $v_t$ and call it trawler critical velocity. This trawler
critical velocity is related to the Landau instability discussed with
the mean-field Gross-Pitaevskii equation in
Ref.\cite{Wu2001PRA,Smerzi2002PRL,Modugno2004PRA}.  Here in this
Letter, this Landau instability is discussed in a more general
setting.

It is easy to demonstrate that the trawler critical velocity is
different from the inside critical velocity.  We consider a limiting
case that the critical velocity is determined by the low energy phonon
excitation (which means that $v_t$ is necessary to be small).  In this
case, we can limit our consideration in the lowest Bloch band, and the
full Hamiltonian of the system can be mapped to an effective one in which
the boson moves in the free space (without the periodic potential) but
with the renormalized dispersion $E(\bf p)$ of 
the lowest Bloch band. For $\bf p \approx
0$, $E({\bf p}) \approx {\bf p}^2 / 2 m^*$ with $m^*$ being the effective
mass.  The effective Hamiltonian in the thermodynamics frame has the form 
\ba
\hat{H}^\prime&=&
\sum_j\frac{(\hat{\bf p}_j+\hbar{\bf k})^2}{2m^*}+\widetilde{H}_{int}\,,
\nonumber\\
&=&\hat{H}_{0}-\frac{m}{m^{*}}{\bf v}\cdot\hat{\bf
  p}+\frac{Nm^*}{2}v^2\,.
\label{eq:eff_ham}
\ea 
where $\hat{H}_{0}$ is the effective Hamiltonian in the superfluid
frame and $\widetilde{H}$ is the interaction potential projected in
the lowest Bloch band. When an elementary excitation with quasi-momentum
$\hbar\bf q$ is generated, its energy is according to Eq.(\ref{eq:eff_ham})
\be \tilde{\varepsilon}^\prime ({\bf v},{\bf q})=\varepsilon_0({\bf q})-
\hbar\frac{m}{m^*}{\bf v}\cdot {\bf q}\,,
\label{eq:latt_excit}
\ee 
where $\varepsilon_0({\bf q})\approx u\hbar{\bf q}$ is the phonon
excitation energy in the superfluid frame.  It is clear that
superfluidity is lost when $v>m^*u_0/m$.  So, the critical velocity is
\be v_t=\frac{m^*}{m}u_0\,,
\label{eq:short_a}
\ee
which is different from the inside critical velocity $v_i$. 
Note that the velocity $m{\bf v}/m^*$ in Eq.(\ref{eq:latt_excit}) 
is the group velocity of the superfluid.

\begin{figure}[!htb]
\includegraphics[width=7.5cm]{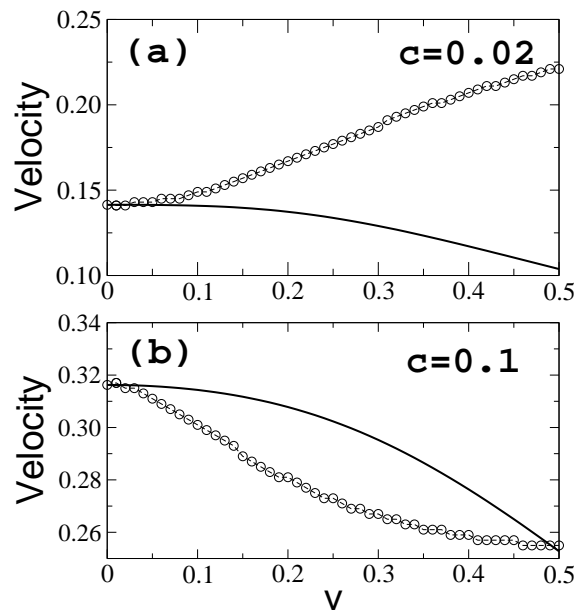}
\caption{Inside critical velocity $v_i$ and trawler critical velocity $v_t$
of a BEC in a one dimensional optical lattice as functions of the lattice
strength $v$. (a) $c=0.02$ (b) $c=0.1$. The solid line is for $v_i$ and the circles with
dashed line represent $v_t$.}
\label{fig:vtvi}
\end{figure}

We now demonstrate these two critical velocities, 
with the simple example,
a BEC in a one-dimensional optical lattice. When this system is
in the superfluid state, it can be well described by the mean-field
Gross-Pitaevskii equation as in Eq.(\ref{eq:gp}). We can
find both critical velocities by numerically computing the Bogoliubov
spectrum of this system as in Ref.\cite{Wu2003NJP}. We have computed two different
cases: in one case the Bloch states corresponding to the trawler
critical velocity are close to the $\Gamma$ point;
in the other these Bloch states are far away from the $\Gamma$ point.
For the first case, the results are  shown in Fig.\ref{fig:vtvi}(a),
where we see these two velocities have different trends: the inside critical
velocities $v_i$ decreases with the lattice strength while the
trawler critical velocity increases. In the other case,
both critical velocities decrease with the lattice strength.

For supersolid helium, the critical velocity
has not been measured up to date. The ``critical velocity''
(3.6-38$\mu$m/s) measured by Kim and Chan\cite{Kim2004Sci}
is likely associated with the vortex quantization in the annular
container. We are aware that the supersolid helium is
a highly controversy topic at present\cite{Ceperley2006NPhys}.

In summary, we have tried to answer a simple question,
``What is the  critical velocity for a superfluid in a periodic potential?''
The answer depends on
the way how the critical velocity is probed. If you probe it
by moving an impurity inside the BEC, you have the inside
critical velocity; if you probe it by moving the lattice, you
obtain the trawler critical velocity. These two velocities
are different in nature and generally in values.
For a BEC in an optical lattice, 
both critical velocities are measurable with current experimental 
techniques\cite{Onofrio2000PRL,Fallani2004PRL}.


We thank A.L. Fetter drawing our attention to the early work on ion motion
in liquid helium. We are supported by the ``BaiRen'' program of the 
Chinese Academy of Sciences and the NSF of China (10504040,10604063) 
B.W is also supported by the 973 project of China (2005CB724500,2006CB921400).


\begin{thebibliography}{23}
\expandafter\ifx\csname natexlab\endcsname\relax\def\natexlab#1{#1}\fi
\expandafter\ifx\csname bibnamefont\endcsname\relax
  \def\bibnamefont#1{#1}\fi
\expandafter\ifx\csname bibfnamefont\endcsname\relax
  \def\bibfnamefont#1{#1}\fi
\expandafter\ifx\csname citenamefont\endcsname\relax
  \def\citenamefont#1{#1}\fi
\expandafter\ifx\csname url\endcsname\relax
  \def\url#1{\texttt{#1}}\fi
\expandafter\ifx\csname urlprefix\endcsname\relax\def\urlprefix{URL }\fi
\providecommand{\bibinfo}[2]{#2}
\providecommand{\eprint}[2][]{\url{#2}}

\bibitem[{\citenamefont{Landau}(1941)}]{Landau1941USSR}
\bibinfo{author}{\bibfnamefont{L.~D.} \bibnamefont{Landau}},
  \bibinfo{journal}{J. Phys. U.S.S.R.} \textbf{\bibinfo{volume}{5}},
  \bibinfo{pages}{71} (\bibinfo{year}{1941}).

\bibitem[{\citenamefont{Landau and Lifshitz}(1981)}]{Landau_SM2}
\bibinfo{author}{\bibfnamefont{L.~D.} \bibnamefont{Landau}} \bibnamefont{and}
  \bibinfo{author}{\bibfnamefont{E.~M.} \bibnamefont{Lifshitz}},
  \emph{\bibinfo{title}{Statistical Physics (II)}}
  (\bibinfo{publisher}{Pergamon, Oxford}, \bibinfo{year}{1981}).

\bibitem[{\citenamefont{Nozieres and Pines}(1990)}]{Nozieres1990Book}
\bibinfo{author}{\bibfnamefont{P.}~\bibnamefont{Nozieres}} \bibnamefont{and}
  \bibinfo{author}{\bibfnamefont{D.}~\bibnamefont{Pines}},
  \emph{\bibinfo{title}{The Theory of Quantum Liquids (II)}}
  (\bibinfo{publisher}{Addison-Wesley, New York}, \bibinfo{year}{1990}).

\bibitem[{\citenamefont{Morsch and Oberthaler}(2006)}]{Morsch2006RMP}
\bibinfo{author}{\bibfnamefont{O.}~\bibnamefont{Morsch}} \bibnamefont{and}
  \bibinfo{author}{\bibfnamefont{M.}~\bibnamefont{Oberthaler}},
  \bibinfo{journal}{Rev. Mod. Phys.} \textbf{\bibinfo{volume}{78}},
  \bibinfo{pages}{179} (\bibinfo{year}{2006}).

\bibitem[{\citenamefont{Kim and Chan}(2004)}]{Kim2004Sci}
\bibinfo{author}{\bibfnamefont{E.}~\bibnamefont{Kim}} \bibnamefont{and}
  \bibinfo{author}{\bibfnamefont{M.~H.~W.} \bibnamefont{Chan}},
  \bibinfo{journal}{Science} \textbf{\bibinfo{volume}{305}},
  \bibinfo{pages}{1941} (\bibinfo{year}{2004}).

\bibitem[{\citenamefont{Andreev and Lifshitz}(1969)}]{Andreev1969JETP}
\bibinfo{author}{\bibfnamefont{A.~F.} \bibnamefont{Andreev}} \bibnamefont{and}
  \bibinfo{author}{\bibfnamefont{I.~M.} \bibnamefont{Lifshitz}},
  \bibinfo{journal}{Sov. Phys. JETP} \textbf{\bibinfo{volume}{29}},
  \bibinfo{pages}{1107} (\bibinfo{year}{1969}).

\bibitem[{\citenamefont{Leggett}(1970)}]{Leggett1970PRL}
\bibinfo{author}{\bibfnamefont{A.~J.} \bibnamefont{Leggett}},
  \bibinfo{journal}{Phys. Rev. Lett.} \textbf{\bibinfo{volume}{25}},
  \bibinfo{pages}{1543} (\bibinfo{year}{1970}).

\bibitem[{\citenamefont{Lattimer and Prakash}(2004)}]{Lattimer2004Sci}
\bibinfo{author}{\bibfnamefont{J.~M.} \bibnamefont{Lattimer}} \bibnamefont{and}
  \bibinfo{author}{\bibfnamefont{M.}~\bibnamefont{Prakash}},
  \bibinfo{journal}{Science} \textbf{\bibinfo{volume}{304}},
  \bibinfo{pages}{536} (\bibinfo{year}{2004}).

\bibitem[{\citenamefont{Carter et~al.}(2005)\citenamefont{Carter, Chamel, and
  Haensel}}]{Carter2005NPA}
\bibinfo{author}{\bibfnamefont{B.}~\bibnamefont{Carter}},
  \bibinfo{author}{\bibfnamefont{N.}~\bibnamefont{Chamel}}, \bibnamefont{and}
  \bibinfo{author}{\bibfnamefont{P.}~\bibnamefont{Haensel}},
  \bibinfo{journal}{Nucl. Phys. A} \textbf{\bibinfo{volume}{759}},
  \bibinfo{pages}{441} (\bibinfo{year}{2005}).

\bibitem[{\citenamefont{Zwierlein et~al.}(2005)\citenamefont{Zwierlein,
  Abo-Shaeer, Schirotzek, Schunck, and Ketterle}}]{Ketterle2005Nature}
\bibinfo{author}{\bibfnamefont{M.~W.} \bibnamefont{Zwierlein}},
  \bibinfo{author}{\bibfnamefont{J.~R.} \bibnamefont{Abo-Shaeer}},
  \bibinfo{author}{\bibfnamefont{A.}~\bibnamefont{Schirotzek}},
  \bibinfo{author}{\bibfnamefont{C.~H.} \bibnamefont{Schunck}},
  \bibnamefont{and} \bibinfo{author}{\bibfnamefont{W.}~\bibnamefont{Ketterle}},
  \bibinfo{journal}{Nature} \textbf{\bibinfo{volume}{435}},
  \bibinfo{pages}{1047} (\bibinfo{year}{2005}).

\bibitem[{\citenamefont{Onofrio et~al.}(2000)\citenamefont{Onofrio, Raman,
  Vogels, Abo-Shaeer, Chikkatur, and Ketterle}}]{Onofrio2000PRL}
\bibinfo{author}{\bibfnamefont{R.}~\bibnamefont{Onofrio}},
  \bibinfo{author}{\bibfnamefont{C.}~\bibnamefont{Raman}},
  \bibinfo{author}{\bibfnamefont{J.~M.} \bibnamefont{Vogels}},
  \bibinfo{author}{\bibfnamefont{J.~R.} \bibnamefont{Abo-Shaeer}},
  \bibinfo{author}{\bibfnamefont{A.~P.} \bibnamefont{Chikkatur}},
  \bibnamefont{and} \bibinfo{author}{\bibfnamefont{W.}~\bibnamefont{Ketterle}},
  \bibinfo{journal}{Phys. Rev. Lett.} \textbf{\bibinfo{volume}{85}},
  \bibinfo{pages}{2228} (\bibinfo{year}{2000}).

\bibitem[{\citenamefont{Fallani et~al.}(2004)\citenamefont{Fallani, Sarlo, Lye,
  Modugno, Saers, Fort, and Inguscio}}]{Fallani2004PRL}
\bibinfo{author}{\bibfnamefont{L.}~\bibnamefont{Fallani}},
  \bibinfo{author}{\bibfnamefont{L.~D.} \bibnamefont{Sarlo}},
  \bibinfo{author}{\bibfnamefont{J.~E.} \bibnamefont{Lye}},
  \bibinfo{author}{\bibfnamefont{M.}~\bibnamefont{Modugno}},
  \bibinfo{author}{\bibfnamefont{R.}~\bibnamefont{Saers}},
  \bibinfo{author}{\bibfnamefont{C.}~\bibnamefont{Fort}}, \bibnamefont{and}
  \bibinfo{author}{\bibfnamefont{M.}~\bibnamefont{Inguscio}},
  \bibinfo{journal}{Phys. Rev. Lett.} \textbf{\bibinfo{volume}{93}},
  \bibinfo{pages}{140406} (\bibinfo{year}{2004}).

\bibitem[{\citenamefont{Pitaevskii}(2003)}]{Pitaevskii2003Book}
\bibinfo{author}{\bibfnamefont{L.~P.} \bibnamefont{Pitaevskii}},
  \emph{\bibinfo{title}{The Physics of Superconductors (I)}}
  (\bibinfo{publisher}{Springer-Verlag}, \bibinfo{year}{2003}),
  chap.~\bibinfo{chapter}{2}.

\bibitem[{\citenamefont{Wu and Niu}(2003)}]{Wu2003NJP}
\bibinfo{author}{\bibfnamefont{B.}~\bibnamefont{Wu}} \bibnamefont{and}
  \bibinfo{author}{\bibfnamefont{Q.}~\bibnamefont{Niu}}, \bibinfo{journal}{New
  J. of Phys.} \textbf{\bibinfo{volume}{5}}, \bibinfo{pages}{104}
  (\bibinfo{year}{2003}).

\bibitem[{\citenamefont{Fetter}(1976)}]{Fetter1976InBook}
\bibinfo{author}{\bibfnamefont{A.~L.} \bibnamefont{Fetter}},
  \emph{\bibinfo{title}{The Physics of Liquid and Solid Helium (I)}}
  (\bibinfo{publisher}{Wiley}, \bibinfo{year}{1976}),
  chap.~\bibinfo{chapter}{3}.

\bibitem[{\citenamefont{Reif and Meyer}(1960)}]{Reif1960PR}
\bibinfo{author}{\bibfnamefont{F.}~\bibnamefont{Reif}} \bibnamefont{and}
  \bibinfo{author}{\bibfnamefont{L.}~\bibnamefont{Meyer}},
  \bibinfo{journal}{Phys. Rev.} \textbf{\bibinfo{volume}{119}},
  \bibinfo{pages}{1164} (\bibinfo{year}{1960}).

\bibitem[{\citenamefont{Girardeau}(1961)}]{Girardeau1961PoF}
\bibinfo{author}{\bibfnamefont{M.}~\bibnamefont{Girardeau}},
  \bibinfo{journal}{Phys. of Fluids} \textbf{\bibinfo{volume}{4}},
  \bibinfo{pages}{279} (\bibinfo{year}{1961}).

\bibitem[{\citenamefont{Choi and Niu}(1999)}]{ChoiAndNiu1999}
\bibinfo{author}{\bibfnamefont{D.-I.} \bibnamefont{Choi}} \bibnamefont{and}
  \bibinfo{author}{\bibfnamefont{Q.}~\bibnamefont{Niu}},
  \bibinfo{journal}{Phys. Rev. Lett.} \textbf{\bibinfo{volume}{82}},
  \bibinfo{pages}{2022} (\bibinfo{year}{1999}).

\bibitem[{\citenamefont{Morsch et~al.}(2001)\citenamefont{Morsch, M\"{u}ller,
  Cristiani, Ciampini, and Arimondo}}]{Morsch2001PRL}
\bibinfo{author}{\bibfnamefont{O.}~\bibnamefont{Morsch}},
  \bibinfo{author}{\bibfnamefont{J.}~\bibnamefont{M\"{u}ller}},
  \bibinfo{author}{\bibfnamefont{M.}~\bibnamefont{Cristiani}},
  \bibinfo{author}{\bibfnamefont{D.}~\bibnamefont{Ciampini}}, \bibnamefont{and}
  \bibinfo{author}{\bibfnamefont{E.}~\bibnamefont{Arimondo}},
  \bibinfo{journal}{Phys. Rev. Lett.} \textbf{\bibinfo{volume}{87}},
  \bibinfo{pages}{140402} (\bibinfo{year}{2001}).

\bibitem[{\citenamefont{Wu and Niu}(2001)}]{Wu2001PRA}
\bibinfo{author}{\bibfnamefont{B.}~\bibnamefont{Wu}} \bibnamefont{and}
  \bibinfo{author}{\bibfnamefont{Q.}~\bibnamefont{Niu}},
  \bibinfo{journal}{Phys. Rev. A} \textbf{\bibinfo{volume}{64}},
  \bibinfo{pages}{061603} (\bibinfo{year}{2001}).

\bibitem[{\citenamefont{Smerzi et~al.}(2002)\citenamefont{Smerzi, Trombettoni,
  Kevrekidis, and Bishop}}]{Smerzi2002PRL}
\bibinfo{author}{\bibfnamefont{A.}~\bibnamefont{Smerzi}},
  \bibinfo{author}{\bibfnamefont{A.}~\bibnamefont{Trombettoni}},
  \bibinfo{author}{\bibfnamefont{P.~G.} \bibnamefont{Kevrekidis}},
  \bibnamefont{and} \bibinfo{author}{\bibfnamefont{A.~R.}
  \bibnamefont{Bishop}}, \bibinfo{journal}{Phys. Rev. Lett.}
  \textbf{\bibinfo{volume}{89}}, \bibinfo{pages}{170402}
  (\bibinfo{year}{2002}).

\bibitem[{\citenamefont{Modugno et~al.}(2004)\citenamefont{Modugno, Tozzo, and
  Dalfovo}}]{Modugno2004PRA}
\bibinfo{author}{\bibfnamefont{M.}~\bibnamefont{Modugno}},
  \bibinfo{author}{\bibfnamefont{C.}~\bibnamefont{Tozzo}}, \bibnamefont{and}
  \bibinfo{author}{\bibfnamefont{F.}~\bibnamefont{Dalfovo}},
  \bibinfo{journal}{Phys. Rev. A} \textbf{\bibinfo{volume}{70}},
  \bibinfo{pages}{043625} (\bibinfo{year}{2004}).

\bibitem[{\citenamefont{Ceperley}(2006)}]{Ceperley2006NPhys}
\bibinfo{author}{\bibfnamefont{D.}~\bibnamefont{Ceperley}},
  \bibinfo{journal}{Nature Phys.} \textbf{\bibinfo{volume}{2}},
  \bibinfo{pages}{659} (\bibinfo{year}{2006}).

\end{thebibliography}

\end{document}